# Autonomous Calibration of MEMS Gyros in Consumer Portable Devices

You Li, Jacques Georgy, *Member, IEEE,* Xiaoji Niu, Qingli Li, and Naser El-Sheimy

*Abstract—* **This paper presents a real-time calibration method for gyro sensors in consumer portable devices. The calibration happens automatically without the need for external equipment or user intervention. Multi-level constraints, including the pseudo-observations, the accelerometer and magnetometer measurements, and the quasi-static attitude updates, are used to make the method reliable and accurate under natural user motions. Walking tests with the Samsung Galaxy S3 and S4 smartphones showed that the method estimate promising calibration results even under challenging motion modes such as dangling and pocket, and in challenging indoor environments with frequent magnetic interferences**.

## I. Introduction

ADVANCES in Micro-Electro-Mechanical Systems (MEMS) technology combined with the miniaturization of electronics have made it possible to produce chip-based sensors, such as inertial sensors (i.e., accelerometers and gyroscopes (gyros)) and magnetometers. MEMS chips are small and lightweight, consumes very little power, and are extremely low-cost [1]. By virtue of these advantages, MEMS sensors have become appropriate candidates for motion tracking and navigation (i.e., determination of attitude, velocity, and position) applications in many electronic devices such as smartphones, gaming systems, toys, and the next generation wearable devices. For consumer portable devices, dead reckoning (DR) is usually the navigation algorithm used to navigate with inertial sensors; thus, the sensors are not dependent on the transmission or reception of signals from an external source [2]. Such self-contained MEMS-based inertial sensors are ideal for providing continuous information for indoor/outdoor navigation [3].

However, low-cost MEMS inertial sensors suffer from significant run-to-run biases and thermal drifts [4]. The performance of typical MEMS sensors that used in smartphones [5] are shown in Table 1. Although lab calibration at room temperature is a useful way to remove many deterministic sensor errors [6], the sensors' readings can be very different due to the restart and the difference between the operational and calibration environments. Also, it is not affordable for the chip manufacturers to conduct thermal calibration of low-cost sensors. Due to the integration process in the inertial navigation mechanization, any sensor errors will accumulate, resulting in increasing navigation errors. For indoor environments, position or velocity updates from GNSS are not always available. If this is the case, the navigation accuracy will degrade faster over time.

TABLE 1. Performance of typical MEMS sensors [5] in smartphones

| Sensor | Error | Range |
|---|---|---|
| Gyros | Initial Biases at 25 °C | ± 5 deg/s |
|  | Variation over -40 to 85 °C | ± 0.24 deg/s/ °C |
| Accelerom-eters | Initial Biases | ± 60 mg |
|  | Variation over -40 to 85 °C | ± 0.64 mg/ °C |

Therefore, a real-time calibration process is needed to mitigate the drift of the inertial sensor errors, especially the gyro biases. The calibration process should happen automatically in the background without the need for user intervention. This is because mainstream consumer and non-professional users should be able to benefit from the calibration and from a better navigation solution using the calibrated sensors without any specific requirement from them. Achieving such a calibration is not easy, especially when working with consumer-grade inertial sensors. Most calibration methods require external equipment or tools to provide a reference for calibration. It is not realistic to expect the users of the electronic products to use a separate tool to calibrate the sensors. Furthermore, current traditional methods to calibrate the inertial sensors without an external tool involve: a) using static periods for gyroscopes calibration when the sensors are fully static; and b) using the gravity vector for accelerometer

Submission Date: December 9, 2014 "This work was supported in part by National Natural Science Foundation of China (41174028, 41304004) and the research funding from Dr. Naser El-Sheimy. The first author (YL) would also thank the China Scholarship Council for its support (201306270139)".

You Li is with the Geomatics Engineering Department, University of Calgary, Calgary, T2N1N4, Canada and GNSS Research Center, Wuhan University, P.R. China (e-mail: liyou331@gmail.com).
Jacques Georgy is with InvenSense Canada, T2L 2K7, Canada (e-mail: jgeorgy@invensense.com).
Xiaoji Niu and Qingli Li are with the GNSS Research Center, Wuhan University, 430079, P.R.China (e-mail: xjniu@whu.edu.cn; qlli@whu.edu.cn). Xiaoji Niu is the corresponding author of this paper.
Naser El-Sheimy is with the Geomatics Engineering Department, University of Calgary, Calgary T2N1N4, Canada, (e-mail: elsheimy@ucalgary.ca).

calibration while ensuring that the IMU covers various attitudes to make sure that the system is observable [7]. The need for static periods limits the scenarios where calibration can happen; the need for various attitudes increases the operation complexity. Furthermore, when the calibration is done automatically in the background, the need for various attitudes will delay having a full calibration without the user involvement to do specific motions.

In this paper, we propose an autonomous calibration method to calculate the gyros in consumer electronics. This method uses a Kalman filter algorithm and utilizes multiple constraints, including the pseudo-observations, the accelerometer and magnetometer measurements, and the quasi-static attitude updates. The advantages of the proposed calibration algorithm includes:

a) The calibration happens automatically without the need for external equipment or user intervention;
b) The algorithm works under natural user motions such as handheld, phoning, dangling, pocket, belt, and backpack. Also, there is no singularity problem when the pitch angle reaches ± 90°;
c) The algorithm works even in indoor environments with frequent magnetic interferences.

This paper is organized as follows. Section 2 reviews the previous relevant works. Section 3 explains the methodology of the calibration algorithm, including the details of multiple constraints. Section 4 shows some results with analysis and Section 5 draws the conclusion.

## II. PREVIOUS WORKS

The commonly used calibration methods include the standard calibration methods [6, 8] and the multi-position calibration methods [7, 9-12]. Standard calibration methods determine sensor errors by comparing the sensor outputs with known reference inputs. Due to the dependence on specialized equipment, the standard methods are always designed for in-lab tests, factory calibration and relatively high-grade IMUs.

To calibrate an IMU just with simple devices or even without any specific tool, multi-position methods are developed. The basic idea of a multi-position method can be stated as follows: the norms of the measured outputs of the accelerometer and gyro cluster are equal to the magnitudes of the given specific force (i.e., gravity) and rotational velocity inputs (i.e., the Earth rotation), respectively [13]. However, the main drawback in using multi-position calibration method is that the gyro reference (the Earth rotation rate) is a weak signal (15 deg/h) which can result in observability problems. Therefore, a single axis turntable is required to provide a strong rotation rate signal [7, 10-12], which limits the multi-position method to laboratories.

To estimate gyro errors without any external equipment, an in-field calibration method has been developed [14]. The accelerometer triad is first calibrated by the multi-position method through multiple quasi-static states generated by hand holding. Then, the outputs from the calibrated accelerometers can be used to calibrate gyros. To avoid the requirement of being static/quasi-static, researches have presented gyro calibration methods such as the vertical gyro (VG) method [15] and the approach that uses accelerometers to estimate the horizontal gyro errors [16]. These methods are efficient in calibrating the horizontal gyros but have limited effect on the vertical gyro [16]. To make all sensor errors observable, user intervention is still required: the user needs to rotate the device to different attitudes to make sure that every gyro axis has the chance to experience the horizontal direction.

In this paper, we remove both the inconvenient user intervention process and the quasi-static assumption by using constraints from multiple sensors and apriori information. The features of the referred previous works and the proposed method are listed in Table 2.

TABLE 2. COMPREHENSIVE CHARACTERISTICS OF THE REFERRED PREVIOUS WORKS AND THE PROPOSED METHOD (IMPROVED BASED ON [16])

| Method and author(s) | Required equipment | Calibration accuracy | Features |
|---|---|---|---|
| Standard methods (e.g. six-position method) | Specialized equipment | High | High precision and reliable |
| Multi-position method *(Lötters et al 1998)* | None (quasi-static required) | Low | Calibrate biases and scale factors of the tri-axial accelerometer under quasi-static conditions without any equipment |
| Improved multi-position methods *(Skog and Händel 2006, Syed et al 2007)* | A single axis angle turntable | High | Calibrate non-orthogonalities of tri-axial accelerometer as well; calibrate gyros using a single axis turntable |
| Improved multi-position method *(Zhang et al 2010)* | A single axis angle turntable | High | Detect the inter-triad misalignment between the accelerometer and gyro triads; Relax the requirement of precise orientation control. |
| Improved multi-position method *(Nieminen et al 2010)* | A single axis rate turntable | High | Exploiting the centripetal accelerations caused by the rotation of the turntable |
| Improved multi-position method *(Fong et al 2008)* | None (quasi-static required) | Low | Calibrate the low-cost gyro triad as well as accelerometers without any equipment |
| In-situ method *(Li et al 2012)* | None (in-situ required) | Low | Introduce pseudo-observations to calibrate the biases and scale factors of gyros and accelerometers without any equipment in a short period (about 30 seconds) |
| Proposed method | None | Low | Using Kalman filter algorithm with multi-level constraints to conduct calibration under natural user motions. |

### III. METHODOLOGY

When compared with previous works, the main advantage of the proposed method is the removing of the requirement of equipment and user intervention. In addition, the proposed method can work in real time under natural human motions in both indoor and outdoor environments. In this section, the details of the proposed method will be given. We will also answer the following questions: a) how can the algorithm work under various human motions? and b) how can the algorithm work in indoor environments with frequent magnetic interferences?

Multi-level constraints are utilized to improve the calibration efficiency and accuracy. The first level is called pseudo-observation updates. This constraint is activated when the change in position between two time epochs is within a limited range. In this case, the constraints $\tilde{r}$ = constant and $\tilde{v}$ = 0 are regarded as an observation of pseudo-position and pseudo-velocity. The uncertainty of position and velocity changes during the calibration process are embodied in the covariance matrix of measurement noise (R) in the Kalman Filter. The R matrix is tuned adaptively according to the IMU outputs. The use of the pseudo-observation updates makes it possible to calculate the gyro errors without external equipment or tools; also, the pseudo-observation updates can be used under natural human motions without special training. Therefore, it is feasible to run the calibration algorithm in the background without user interaction.

The second level of constraints is from accelerometers and magnetometers. We use the accelerometer and magnetometer measurements in a tightly-coupled way, which brings several benefits for the pedestrian navigation applications with various phone displacements. The details will be introduced later in this section.

Another advantage of our method is that it maximizes the contribution from magnetometers in environments with frequent magnetic perturbations. We make the magnetometer measurements reliable based on the following fact: to aid gyro calibration, what we need is attitude changes, instead of the absolute attitude. The proposed method can use the magnetometer measurements without knowing the absolute values of the local magnetic field (LMF) parameters (i.e., declination, inclination, and magnitude).

Moreover, the proposed calibration method is different from the traditional method which adds sensor errors into the navigation Kalman filter. The main objective of the navigation algorithm is to estimate the navigation states (i.e., position, velocity, and attitude) instead of sensor errors. Therefore, when using extra apriori information or setting parameters, it is preferable to assure that any inaccurate estimate of sensor errors will not destroy the navigation algorithm rather than to estimate residual sensor errors with a higher accuracy.

However, in the proposed calibration method, we use specific updates, such as the pseudo-observations, and set the Kalman filter parameters with the aim of maximizing the calibration accuracy. Also, the calibration results can be evaluated before feedback to avoid the degradation of the whole navigation system under extreme navigation conditions.

The algorithm is comprised of the IMU sensor error models, the INS mechanization, and the Kalman filter models. The INS mechanization follows [17] and will not be described in detail. For details about Kalman filter the reader can refer to [18]. The following sub-sections will introduce the algorithm, including system error models, the system model and the measurement model (for updates from multiple sensors and apriori information).

#### A. Sensor Error Models

A major problem of applying MEMS sensors is the changes of the biases and the scale factors [4]. With the current MEMS manufacturing technology, non-orthogonality errors are relatively smaller compared to the biases and scale factor errors. Therefore, only biases and scale factor errors are taken into account. The output error equations of accelerometers and gyros can be described respectively as below:

$$\delta \mathbf{f}^b = \mathbf{b}_a + diag(\tilde{\mathbf{f}}^b)\delta \mathbf{s}_a + \mathbf{w}_a \quad (1)$$

$$\delta \boldsymbol{\omega}_{ib}^b = \mathbf{b}_g + diag(\tilde{\boldsymbol{\omega}}_{ib}^b)\delta \mathbf{s}_g + \mathbf{w}_g \quad (2)$$

where $\delta \mathbf{f}^b$ and $\delta \boldsymbol{\omega}_{ib}^b$ are the error vectors of specific force and angular rate, respectively. $\mathbf{b}_a$ and $\mathbf{b}_g$ are the accelerometer and gyro biases. $\delta \mathbf{s}_a$ and $\delta \mathbf{s}_g$ are the linear scale factor error vectors, $\mathbf{w}_a$ and $\mathbf{w}_g$ represent the sensor noises, and $\tilde{\mathbf{f}}^b$ and $\tilde{\boldsymbol{\omega}}_{ib}^b$ are the measured specific force and angular rate, respectively. The symbol $diag(\cdot)$ indicates the diagonal matrix form of a vector.

The sensor biases and scale factor errors are modeled as first-order Gauss-Markov processes [19]. Take the gyro biases as an example:

$$\dot{\mathbf{b}}_g = -(1/\tau_{bg})\mathbf{b}_g + \mathbf{w}_{bg} \quad (3)$$

where $\tau_{bg}$ denotes for the correlation time of the gyro biases and $\mathbf{w}_{bg}$ is the driving noise vector.

#### B. Kalman Filter - System Model

A simplified form of the psi-angle error model [17] is applied as the continuous-time state equations in the Kalman filter [16].

$$\begin{bmatrix} \delta \dot{\mathbf{r}}^n \\ \delta \dot{\mathbf{v}}^n \\ \dot{\boldsymbol{\psi}} \end{bmatrix} = \begin{bmatrix} -\boldsymbol{\omega}_{en}^n \times \delta \mathbf{r}^n + \delta \mathbf{v}^n \\ -(2\boldsymbol{\omega}_{ie}^n + \boldsymbol{\omega}_{en}^n) \times \delta \mathbf{v}^n + \mathbf{f}^n \times \boldsymbol{\psi} + \mathbf{C}_b^n \delta \mathbf{f}^b \\ -(\boldsymbol{\omega}_{ie}^n + \boldsymbol{\omega}_{ec}^n) \times \boldsymbol{\psi} - \mathbf{C}_b^n \delta \boldsymbol{\omega}_{ib}^b \end{bmatrix} \quad (4)$$

where $\delta \mathbf{r}^n$, $\delta \mathbf{v}^n$ and $\boldsymbol{\psi}$ are the errors of position, velocity, and attitude. $\mathbf{C}_b^n$ is the Direction Cosine Matrix (DCM) from b-frame (i.e., the body frame) to n-frame (i.e., the navigation frame). $\mathbf{f}^n$ is the specific force vector projected to n-frame, and $\boldsymbol{\omega}_{ie}^n$ and $\boldsymbol{\omega}_{en}^n$ represent the angular rate of the Earth and that of n-frame with respect to e-frame (i.e. the Earth frame), both projected to n-frame. The symbol "×" denotes cross product of two vectors. $\delta \mathbf{f}^b$ and $\delta \boldsymbol{\omega}_{ib}^b$ are the output errors of accelerometers and gyros, as explained in (1) and (2).

*C. Kalman Filter - Measurement Model*

Different kinds of constraints are used to build the measurement model, including the pseudo-observations, the accelerometer and magnetometer measurements, and the quasi-static attitude updates.

*1) Pseudo-observations*

The pseudo-position and pseudo-velocity observations are proposed based on the fact that the range of the position and linear velocity of the IMU are within a limited scope [16]. In this paper, we use the pseudo-position update while walking with natural human motions. The measurement model of pseudo-position is

$$\hat{\mathbf{r}}^n - \tilde{\mathbf{r}}^n = \delta \mathbf{r}^n + \mathbf{n}_1 \quad (5)$$

with $\tilde{\mathbf{r}}^n = \mathbf{constant}$

where $\hat{\mathbf{r}}^n$ and $\tilde{\mathbf{r}}^n$ are the position vectors from the INS mechanization and pseudo-position, respectively; $\delta \mathbf{r}^n$ is the position errors vector; and $\mathbf{n}_1$ is the measurement noises (i.e. the inaccuracy) of the pseudo-position.

Here is a feasible way of setting the $\mathbf{R}$ matrix for the pseudo-position: first a set of initial position noises are set roughly; then the position changes during a period can be calculated; based on this result, the elements in $\mathbf{R}$ could be further tuned. This process is done autonomously by the software.

*2) Accelerometer measurement model*

In this paper, the purpose of using accelerometers and magnetometers focuses on providing most accurate gyro bias estimation, which is different from the attitude and heading reference systems (AHRS) [20]. We build the measurement model by using the accelerometer readings directly, instead of using the accelerometers-derived roll and pitch angles. This is important for the pedestrian navigation applications with arbitrary phone displacements, since it avoids the singularity problem when the pitch angle reaches ± 90°. The tightly-coupled accelerometer measurement model is [21]

$$\delta \mathbf{f}^n = \mathbf{f}^n - \hat{\mathbf{f}}^n = \mathbf{f}^n - \hat{\mathbf{C}}_b^n \tilde{\mathbf{f}}^b \quad (6)$$

When neglecting the accelerometer deterministic errors,

$$\begin{aligned} \delta \mathbf{f}^n &= \mathbf{f}^n - \left(\mathbf{I} - [\boldsymbol{\psi} \times]\right) \mathbf{C}_b^n \mathbf{f}^b + \hat{\mathbf{C}}_b^n \mathbf{n}_2 \\ &= -[\boldsymbol{\psi} \times] \mathbf{g}^n + \hat{\mathbf{C}}_b^n \mathbf{n}_2 \\ &= [\mathbf{g}^n \times] \boldsymbol{\psi} + \hat{\mathbf{C}}_b^n \mathbf{n}_2 \end{aligned} \quad (7)$$

where, $\mathbf{f}^n = -\mathbf{g}^n = \begin{bmatrix} 0 & 0 & -g \end{bmatrix}^T$, $\hat{\mathbf{C}}_n^b$ is the DCM provided by the Kalman filter, $g$ is the local gravity value, $\boldsymbol{\psi}$ is the attitude error, and $\mathbf{n}_2$ is the noise.

For pedestrian applications, the acceleration are commonly high-frequency and alternating. Thus, it is reasonable to model the actual accelerations as measurement noises. The components in $\mathbf{R}$ related to the accelerometer measurements are set based on the value of the actual linear acceleration $A$.

$$A = \left| norm(\mathbf{f}^b) - g \right| \quad (8)$$

When $A \leq |Th_{acc1}|$ (i.e., in non-acceleration mode), the corresponding components in $\mathbf{R}$ are set as $\sigma_a^2$, where $\sigma_a^2$ is set according to the specifications of the accelerometer used.

When $|Th_{acc1}| \leq A \leq |Th_{acc2}|$ (i.e., in low-acceleration mode), the acceleration uncertainties is set as $s(A^2 / P)\sigma_a^2$, where $P$ is the corresponding components of attitudes in the covariance matrix and $s$ is a scalar. This parameter setting method follows the research in [21].

When $A \geq |Th_{acc2}|$ (i.e., in high-acceleration mode), the accelerometer is far away from the truth. Accordingly, the components in $\mathbf{R}$ are set as a large number $\sigma_{aMax}^2$. In this situation, the accelerometer measurements will not contribute to the solution.

*3) Magnetometer measurement model*

The perturbations in the magnetometer measurements are different from that in the accelerometers. The latter is commonly high-frequency and alternating; however, the magnetic perturbations are caused by external magnetic bodies such as man-made infrastructures. Therefore, a typical type of magnetic perturbation is that both the direction and strength of the LMF are changed, but the change is stable within a limited space (or periods). The period during which the LMF is stable can be called as quasi-static magnetic field (QSMF) period, and can be detected by using the magnitude of magnetometer readings [22].

In this paper, we use magnetometer measurements to improve the gyro calibration during QSMF periods. It is assumed that we have totally no idea about the LMF parameters. The details about these magnetic field parameters can refer to [23]. Instead, we calibrate the LMF at the beginning of each QSMF period. The flowchart of using magnetometers under QSMF periods is shown in Figure 1.

The LMF vector during the k-th QSMF period is calibrated by:

$$\mathbf{m}_k^n = (\hat{\mathbf{C}}_n^b)^T \tilde{\mathbf{m}}_{k,1}^b \qquad (9)$$

where $\tilde{\mathbf{m}}_{k,1}^b$ is the magnetometer reading at the beginning of the first epoch(s) of k-th QSMF period. The computed $\mathbf{m}_k^n$ is then used as the reference during the k-th QSMF period.

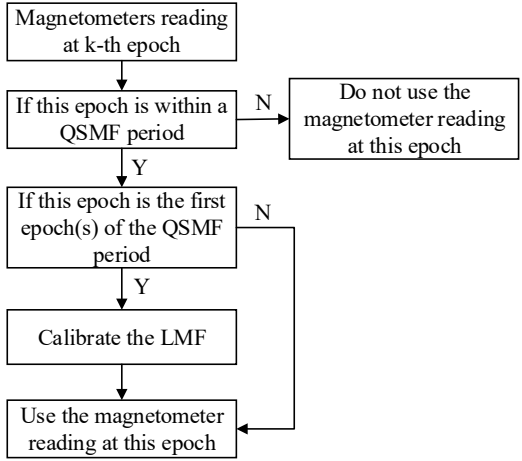

Figure 1. The flowchart of using magnetometer data during QSMF periods

The measurement model is built by using the magnetometer readings directly, which avoid the leveling (i.e., using the accelerometer readings to calculate the roll and pitch angles) step. Therefore, the measurement model, as shown in (10), is independent from the accelerometer measurements.

$$\delta \mathbf{m}^n = [\mathbf{m}^n \times] \boldsymbol{\psi} + \hat{\mathbf{C}}_b^n \mathbf{n}_3 \qquad (10)$$

where, $\delta \mathbf{m}^n = \hat{\mathbf{C}}_b^n \tilde{\mathbf{m}}^b - \mathbf{m}^n$, $\tilde{\mathbf{m}}^b$ is the magnetometer measurement, $\mathbf{m}^n$ is the calibrated LMF vector, $\boldsymbol{\psi}$ is the attitude error, and $\mathbf{n}_3$ is the measurement noise.

*4) Quasi-static attitude updates*

The assumption for this constraint is that any rotation sensed by the gyros should be caused by the gyro biases when the device is quasi-static. Therefore, it is feasible to improve the gyro calibration during not only strict static periods, but also any quasi-static periods such as the periods when the user stands in-situ and the phone is handheld, phoning, or putting in pocket. The detection of quasi-static periods has been detailed in [24]. The quasi-static attitude updates (QSAU) can be written as

$$\tilde{\boldsymbol{\omega}}_{ib}^b = \mathbf{b}_g + \mathbf{n}_4 \qquad (11)$$

where $\tilde{\boldsymbol{\omega}}_{ib}^b$ is the output vector of a quasi-static gyro triad, $\mathbf{b}_g$ is the gyro biases, and $\mathbf{n}_4$ is the noise.

*D. Parameter setting and initialization*

The Kalman filter parameters include the initial state vector $\mathbf{x}_0$ (i.e., the vector including the error states $\delta \mathbf{r}$, $\delta \mathbf{v}$, $\boldsymbol{\psi}$, $\mathbf{b}_a$, $\mathbf{b}_g$, $\delta \mathbf{s}_a$, and $\delta \mathbf{s}_g$) and the corresponding initial values of the INS mechanization (i.e., the initial values of position $\hat{\mathbf{r}}(t_0)$, velocity $\hat{\mathbf{v}}(t_0)$, and attitude $\hat{\mathbf{C}}_b^n(t_0)$), the initial covariance matrix of state vector ($\mathbf{P}_0$), the covariance matrix of system noise ($\mathbf{Q}$) and the covariance matrix of measurement noise ($\mathbf{R}$).

The parameters can be set as $\mathbf{x}_0 = \mathbf{0}$ (i.e., $\delta \mathbf{r} = \delta \mathbf{v} = \boldsymbol{\psi} = \mathbf{b}_a = \mathbf{b}_g = \delta \mathbf{s}_a = \delta \mathbf{s}_g = \mathbf{0}$), $\hat{\mathbf{r}}(t_0) = \mathbf{r}_0$, and $\hat{\mathbf{v}}(t_0) = \mathbf{0}$, where $\mathbf{r}_0$ is the approximate IMU position. The initial DCM $\hat{\mathbf{C}}_b^n(t_0)$ can be determined by [24]

$$\tilde{\mathbf{C}}_b^n = \left( \begin{bmatrix} \mathbf{f}^n & \mathbf{m}^n & \mathbf{l}^n \end{bmatrix}^T \right)^{-1} \begin{bmatrix} \tilde{\mathbf{f}}^b & \tilde{\mathbf{m}}^b & \tilde{\mathbf{l}}^b \end{bmatrix}^T \qquad (12)$$

where $\mathbf{f}^n$ and $\mathbf{m}^n$ are the specific force and LMF vector in the navigation frame, $\mathbf{l}^n = \mathbf{f}^n \times \mathbf{m}^n$; $\tilde{\mathbf{f}}^b$ and $\tilde{\mathbf{m}}^b$ are the accelerometer and magnetometer measurements, and $\tilde{\mathbf{l}}^b = \tilde{\mathbf{f}}^b \times \tilde{\mathbf{m}}^b$. If the LMF is not quasi-static, the roll and pitch angles are calculated from the accelerometer measurement, and the heading angle is set as zero.

The proposed method will be tested with different smartphones under natural human motions.

## IV. TESTS AND RESULTS

Different outdoor and indoor walking tests were conducted with three smartphones. The tested motion modes comprised typical phone locations and attitudes including handheld, phoning, dangling, in pocket, in belt, and in backpack. At the end of each test, there was a quasi-static period to calculate the reference values of gyro biases. The tested motion modes and the corresponding gyro and accelerometer readings are shown in Figure 2. We can see that dangling and pocket have the strongest gyro dynamics.

The tests were performed with Samsung Galaxy S3 and S4 smartphones. To make the gyro errors more significant (i.e., to test the calibration algorithm), gyro biases of 3 deg/s, -3 deg/s and 3 deg/s were added into the raw gyro outputs before data processing. The rough reference values of the gyro biases are shown in Table 2.

TABLE 2. ROUGH REFERENCE VALUES OF GYRO BIASES

|  | Phone #1 | Phone #2 | Phone #3 |
|---|---|---|---|
| Gyro X (deg/s) | 0.9 | 6.2 | 1.3 |
| Gyro Y (deg/s) | -2.1 | -2.1 | -2.6 |
| Gyro Z (deg/s) | 2.8 | 4.0 | 2.0 |

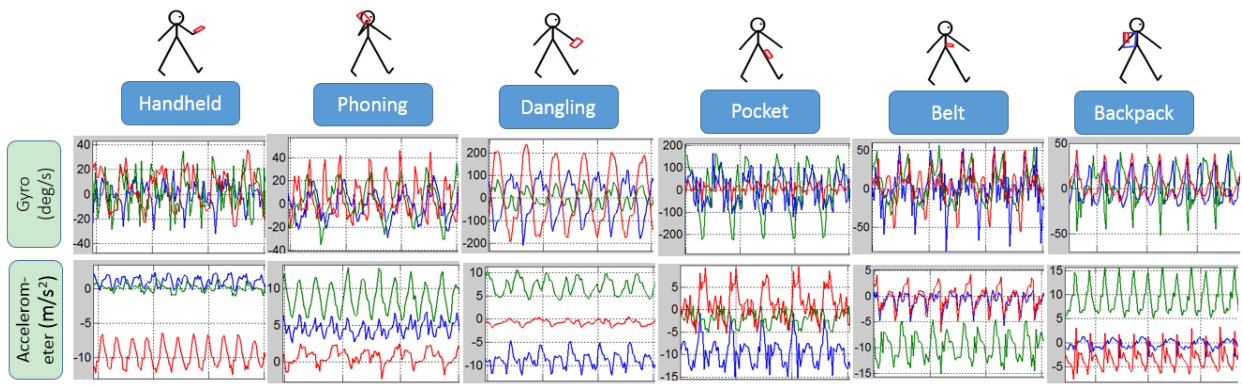

Figure 2. Tested motion modes and corresponding gyro and accelerometer readings

## A. Difference between outdoors and indoors

The main difference between indoor and outdoor environments is the existence of magnetic perturbations. The commonly used method of using magnetometers is straightforward and consist of three steps [25]: a) leveling the magnetometer readings by using the accelerometers; b) calculating the magnetic heading by using the leveled magnetometers; and c) calculating the true heading by a declination angle to the magnetic heading. This is based on the assumption that the LMF is simply the geomagnetic field and thus declination angle can be calculated from the IGRF model [26]. In real tests, we found that the majority of the outdoor tests met this assumption; however, in the indoor tests, the LMF could easily be different from the geomagnetic field, and might vary from point to point.

Table 3 provide a sample of the outdoor and indoor magnetic environments. All figures are plotted using the magnetic information while walking (handheld) case. The figures in the second row of the table show the magnetometer readings and their magnitudes (cyan lines). The yellow dots indicate the QSMF periods. The figures in the last row show the calibrated LMF. The LMF kept stable during this outdoor test; on the other hand, it varied significantly during the indoor test.

The outdoor and indoor tests will be given separately in the following two subsections.

TABLE 3. MAGNETOMETER READINGS AND CALIBRATED LMF WHEN WALKING (HANDHELD) IN OUTDOORS AND INDOORS

| | Outdoor | Indoor |
|---|---|---|
| Magnetometer readings | Mag readings plot | Mag readings plot |
| Calibrated LMF | Calibrated Magnetic Reference for each QSMF | Calibrated Magnetic Reference for each QSMF |

## B. Outdoor walking tests

The outdoor test environment and trajectory are shown in Figure 3. It is a sidewalk crossing different parking lots. Thus, there is no buildings within 10 meters from the sidewalk.

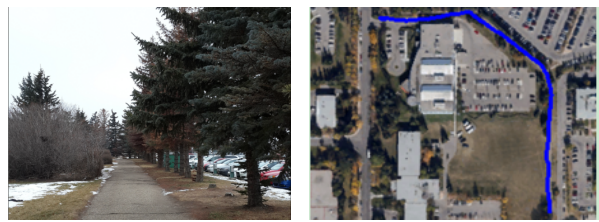

Figure 3. Outdoor test environment and trajectory

The figures of the calibraiton results are shown in Table 4.

The first column indicates the tested motion modes, while the other three columns are the results of three phones. In each plot, the magenta dots indicate the availability of the magnetometer measurements. The pseudo-observation and accelerometer measurements were always available during these tests; therefore, the indicators for these constraints are not shown.

TABLE 4. FIGURES OF CALIBRATION RESULTS IN OUTDOOR TESTS (MAGENTA DOTS INDICATE THE AVAILABILITY OF MAGNETOMETER MEASUREMENTS)

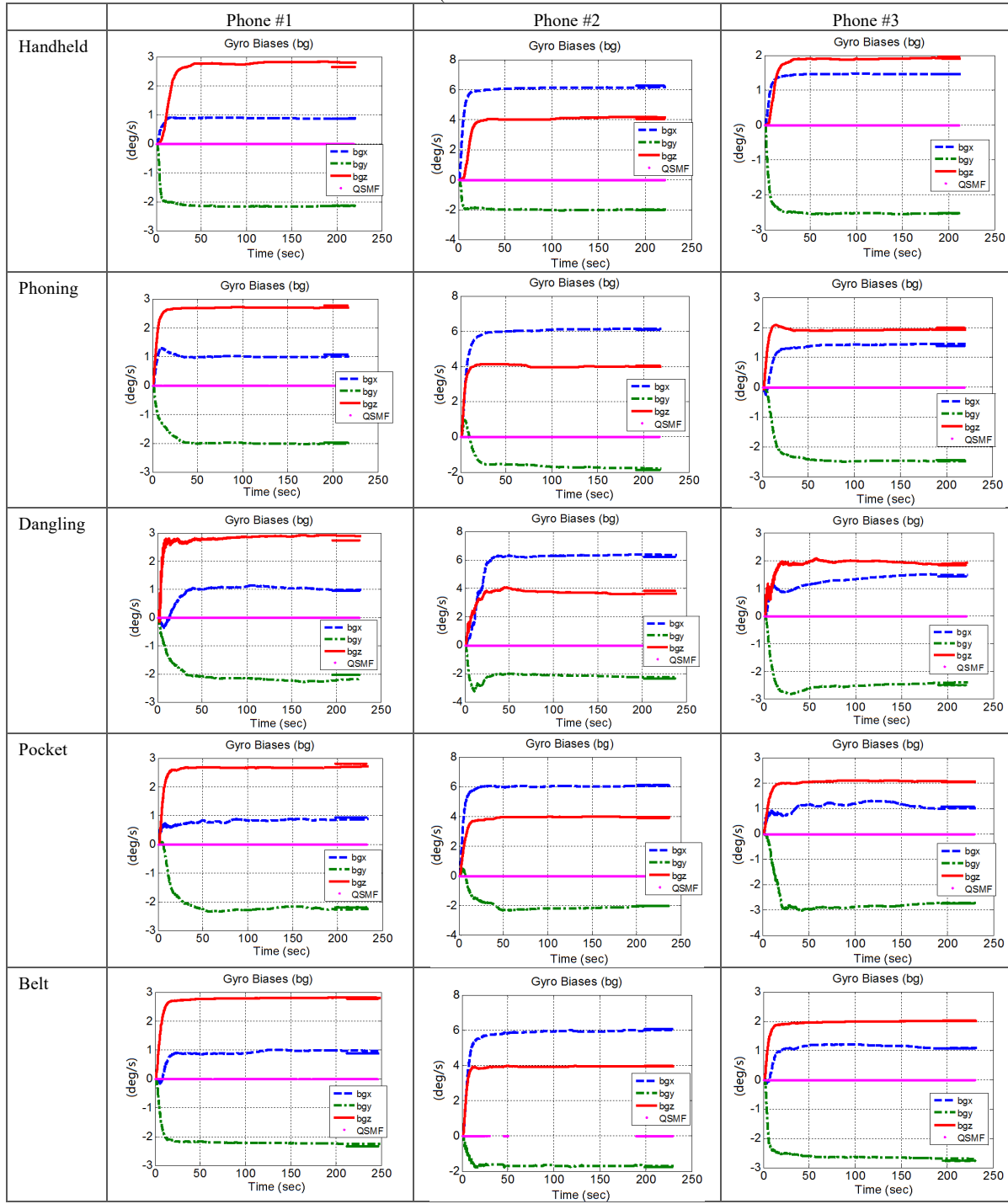

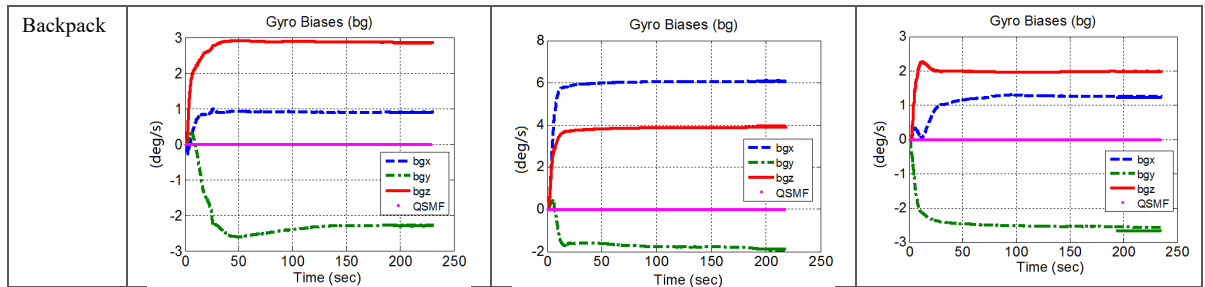

Most of the calibrated sensor errors have converged in the first 30 s, and all of them converged within 50 s. Table 5 shows the statistical results of the calibration errors under all scenarios with different phones. The mean and RMS errors were calculated using the differences between the calibration results at every epoch after convergence and the reference values obtained by averaging the gyro readings during the quasi-static periods at the end of each test.

TABLE 5. STATISTICAL RESULTS OF THE CALIBRATION ERRORS UNDER ALL SCENARIOS

| Phone | Values (deg/s) | X | Y | Z |
|---|---|---|---|---|
| #1 | Reference true value | 0.971 | -2.099 | 2.776 |
|  | Mean error | 0.0588 | 0.0645 | 0.0845 |
|  | RMS error | 0.0682 | 0.0843 | 0.0986 |
| #2 | Reference | 6.173 | -2.017 | 3.981 |
|  | Mean | 0.0737 | 0.0655 | 0.0743 |
|  | RMS | 0.0787 | 0.0752 | 0.0883 |
| #3 | Reference | 1.291 | -2.587 | 1.973 |
|  | Mean | 0.0603 | 0.0583 | 0.0503 |
|  | RMS | 0.0707 | 0.0645 | 0.0605 |

The gyro biases reduced from several deg/s to under 0.1 deg/s. Although the gyros within different phones have different biases and some are more significant (e.g., 6 deg/s in phone #2), the results are all at the same level. This indicates the possible accuracy of the calibration method.

To investigate the effect of human motions on the calibration, we also calculated the statistical results under different motion modes. The results are shown in Table 6.

TABLE 6. STATISTICAL RESULTS OF THE CALIBRATION ERRORS UNDER ALL SCENARIOS

| Motions | RMS error (deg/s) | | |
|---|---|---|---|
|  | X | Y | Z |
| Handheld | 0.0633 | 0.0181 | 0.0589 |
| Phoning | 0.0610 | 0.0563 | 0.0643 |
| Dangling | 0.1157 | 0.1144 | 0.1230 |
| Pocket | 0.0882 | 0.0761 | 0.0815 |
| Belt | 0.0675 | 0.0666 | 0.0266 |
| Backpack | 0.0379 | 0.0682 | 0.0412 |

Dangling and pocket have larger calibration errors than the other motions modes. This meets our expectation, since both dangling and pocket provide stronger smartphone dynamics, as shown in Figure 2. Even under such challenging conditions, the gyro biases were reduced to under 0.13 deg/s and 0.1 deg/s, respectively.

### C. Indoor walking tests

The indoor tests were conducted at the main floor of the Energy Environment Experiential Learning (EEEL) building at the University of Calgary, which has a size of approximate 120 × 40 m$^2$. EEEL is a relatively new building with well-equipped facilities. Accordingly, the magnetic perturbations are also significant in this building, which makes it an appropriate place for the indoor tests. The test environment and trajectory are shown in Figure 3.

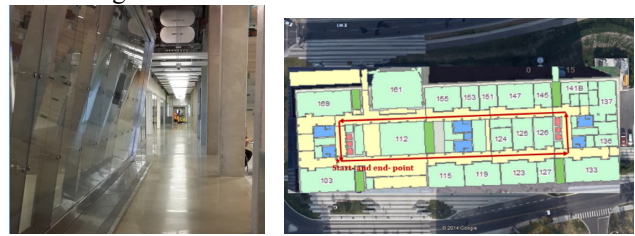

Figure 4. Indoor test environment trajectory

The result figures are shown in Table 7. The solid line at the end of each curve indicates the reference value calculated from the quasi-static data.

TABLE 7. FIGURES OF CALIBRATION RESULTS IN INDOOR TESTS (MAGENTA DOTS INDICATE THE AVAILABILITY OF MAGNETOMETER MEASUREMENTS)

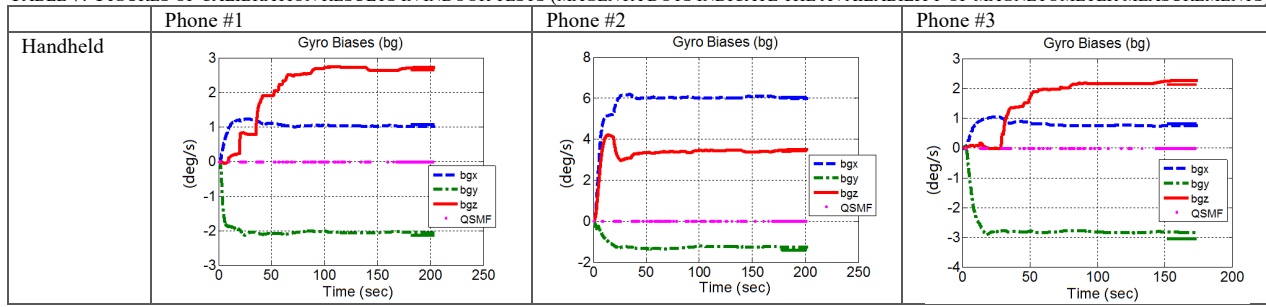

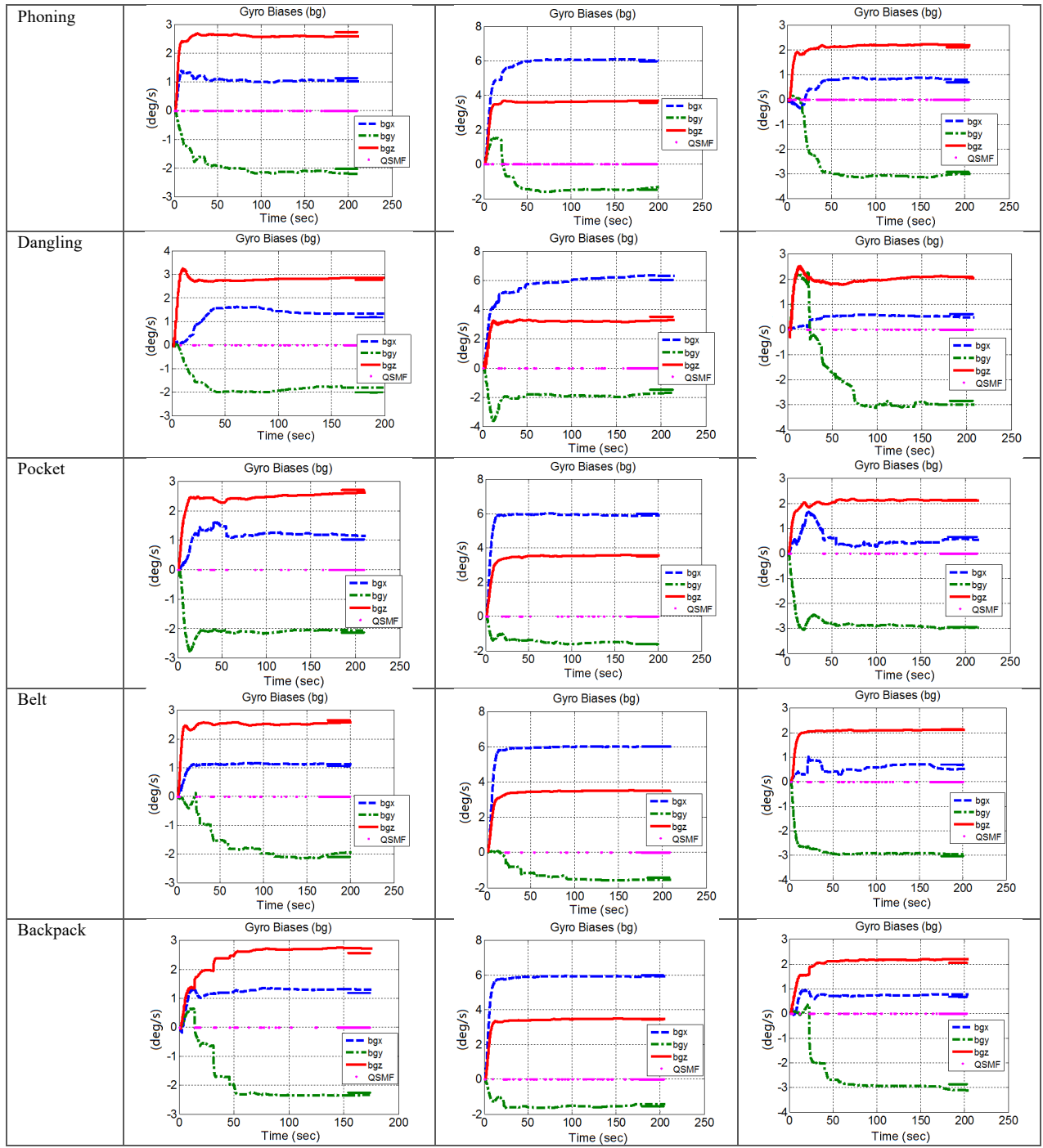

The discontinuity of the periods with magnetometer updates indicates the frequent magnetic perturbations indoors. The convergence of gyro biases is not as smooth as those outdoors; however, even under such challenging environments, all the gyro biases have converged to the right values within 100 s under different human motions. This has verified the feasibility of the proposed method for pedestrian applications. Table 8 shows the statistical results of calibration errors under all scenarios with different phones, and Table 9 classifies the results by motion modes.

| Phone | Values (deg/s) | X | Y | Z |
|---|---|---|---|---|
| #1 | Reference | 1.175 | -2.226 | 2.648 |
| | Mean | 0.0825 | 0.0978 | 0.1032 |
| | RMS | 0.0886 | 0.1122 | 0.1049 |
| #2 | Reference | 6.101 | -1.356 | 3.562 |
| | Mean | 0.0637 | 0.1272 | 0.0550 |
| | RMS | 0.0772 | 0.1298 | 0.0666 |
| #3 | Reference | 0.826 | -3.043 | 2.129 |
| | Mean | 0.1183 | 0.1140 | 0.0847 |
| | RMS | 0.1270 | 0.1157 | 0.1063 |

TABLE 8. STATISTICAL RESULTS OF THE CALIBRATION ERRORS UNDER ALL SCENARIOS

TABLE 9. STATISTICAL RESULTS OF THE CALIBRATION ERRORS UNDER ALL SCENARIOS

| Motions | RMS (deg/s) | | |
|---|---|---|---|
| | X | Y | Z |
| Handheld | 0.0534 | 0.1105 | 0.0786 |
| Phoning | 0.0930 | 0.1134 | 0.0920 |
| Dangling | 0.1376 | 0.1684 | 0.1462 |
| Pocket | 0.1364 | 0.1051 | 0.0789 |
| Belt | 0.0774 | 0.1052 | 0.0707 |
| Backpack | 0.0695 | 0.1031 | 0.0792 |

Table 8 indicates that the proposed calibration method reduced the gyro biases from several deg/s to under 0.13 deg/s in the indoor tests, which is larger than the 0.1 deg/s outdoors. These results are promising for MEMS sensors, since the indoor environment is much harsher the outdoor. In Table 9, the largest calibration errors under dangling and pocket are 0.17 and 0.14 deg/s, which are still larger than those under other motions. The calibration errors under handheld, phoning, belt, and backpack are less than 0.12 deg/s.

Comparing with previous methods such as the vertical gyro method or the methods that use accelerometers to estimate the horizontal gyro errors, an advantage of the proposed method is to use magnetometers during QSMF periods to calibrate the vertical component of gyro biases. Natural human motion signals are usually periodic, as indicated in Figure 2; therefore, not all the gyro axis has the chance to move to the horizontal direction. Thus, the magnetometers measurements during QSMF periods is important. Table 10 show the calibration results using magnetometers measurements under QSMF and that totally ignoring magnetometers. These results further verify the effectiveness of the proposed method in calibrating the vertical gyro bias component.

TABLE 10. FIGURES OF CALIBRATION RESULTS USING MAGNETOMETERS UNDER QSMF AND THAT TOTALLY IGNORING MAGNETOMETERS (WITH PHONE #1)

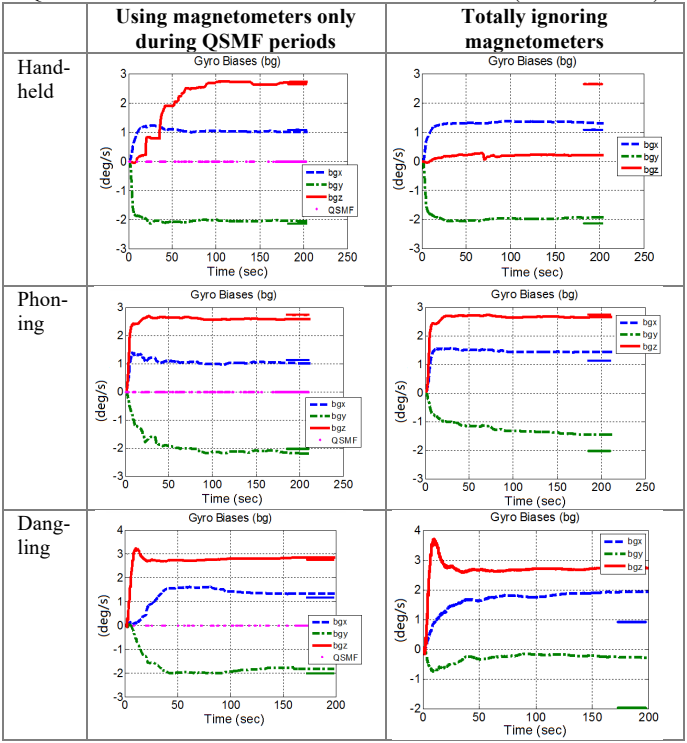

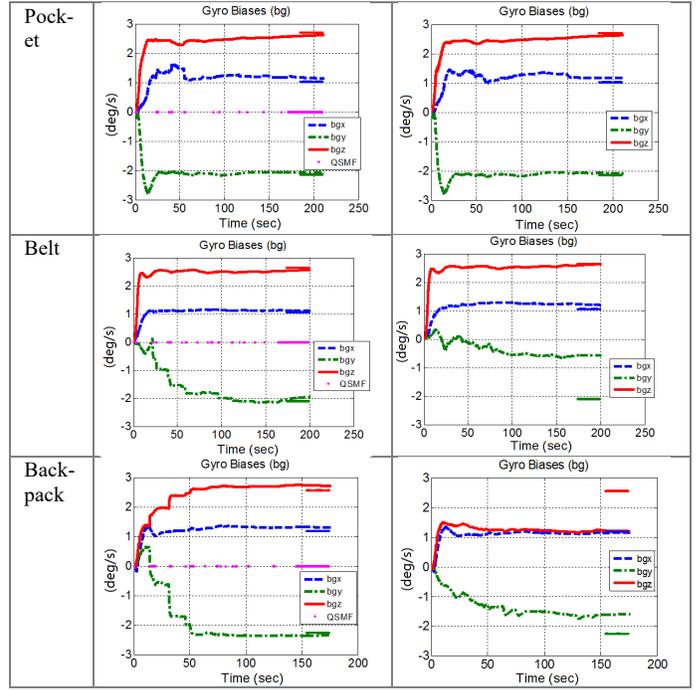

### D. Summary of test results

To make a summary, Figure 5 shows the gyro biases without and with the proposed real-time calibration. The gyro biases of tested phones were reduced from several deg/s to less than 0.15 even in indoor environments. Although the tested phones have different gyro bias values, the calibration errors are at the same level.

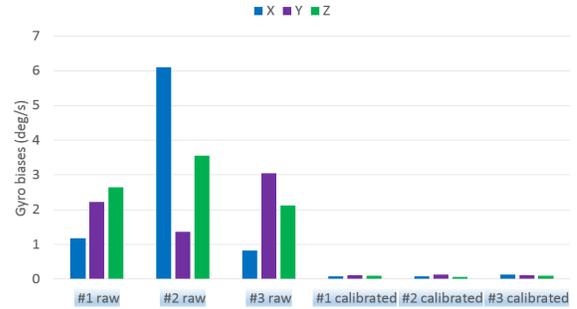

Figure 5. Gyro biases without and with the proposed real-time calibration

Figure 6 compares the calibration errors under different motion modes both indoors and outdoors. The results are better outdoors than indoors under all motion modes. This is most probably because of the harsh magnetic environment indoors. Dangling and pocket are two challenging motion modes for gyro calibration. Their calibration errors are 0.17 and 0.14 deg/s indoors, and 0.13 and 0.09 deg/s outdoors. Under other motions, i.e., handheld, phoning, belt, and backpack, the calibration errors are under 0.12 deg/s indoors and 0.07 outdoors. The calibration results are promising for low cost MEMS sensors in consumer portable devices, even when considering that a part of the calibration errors may be caused by the temperature variations of the gyro errors themselves.

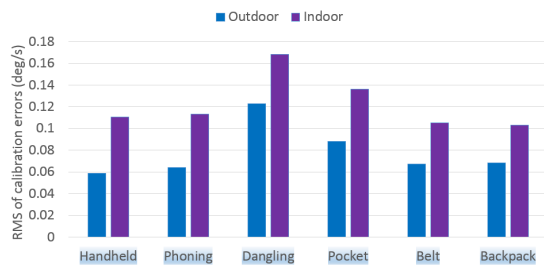

Figure 6. Calibration errors under typical motions both indoors and outdoors

## V. CONCLUSIONS AND FUTURE WORKS

This paper proposed a real-time calibration method without external equipment and without user intervention for gyro sensors in portable devices. The method was tested under walking tests with typical human motions, including handheld, phoning, dangling, pocket, belt, and backpack, both outdoors and indoors. The gyro biases of tested smartphones were reduced from several deg/s to under 0.15 deg/s indoors and 0.1 deg/s outdoors. Under the most challenging motion modes for sensors-based pedestrian navigation, i.e., dangling and pocket, the calibration errors were 0.17 and 0.14 deg/s indoors, and 0.13 and 0.09 deg/s outdoors. Under other motions, the calibration errors are less than 0.12 deg/s indoors and 0.07 outdoors. This calibration method can work in real-time and has a potential for calibration of the MEMS gyros within consumer electronics.

Future works will focus on optimizing the proposed method, e.g., reducing the computational load and further improving the convergence rate.

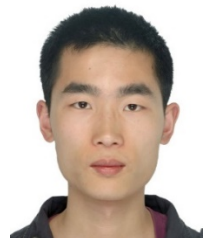

**You Li** is a PhD candidate at both the MMSS Research Group at the University of Calgary and the GNSS Research Center at Wuhan University. He was a software algorithm designer at InvenSense Canada. You Li's research interests are MEMS sensors, the multi-sensor integration technologies, and their applications. He has published 15 papers in referred journals and conferences, and 3 patents.

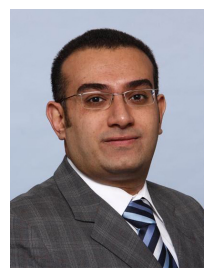

**Jacques Georgy** (S'09–M'10) is the Director of Navigation R&D at InvenSense Inc. He was the VP of R&D and a co-founder of Trusted Positioning Inc. until it was acquired by InvenSense Inc.. He received his Ph.D. degree in Electrical and Computer Engineering from Queen's University, Canada in 2010, and his B.Sc. and M.Sc. degrees in Computer and Systems Engineering from Ain Shams University, Egypt, in 2001 and 2007, respectively. He is working in positioning and navigation systems for portable, vehicular, and machinery applications. His research interests include linear and nonlinear state estimation, positioning and navigation systems, autonomous mobile robot navigation, and underwater target tracking. He has 1 issued patent, 26 distinct patents pending, has co-authored a book, and authored or co-authored over 70 papers. He was the recipient of the Institute of Navigation's 2013 Early Achievement Award for contributions to portable and indoor navigation using MEMS inertial sensors on consumer devices.

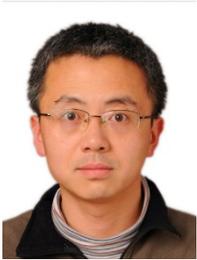

**Xiaoji Niu** is a Professor of GNSS Research Center at Wuhan University in China. He got his Ph.D. and bachelor degrees from the Department of Precision Instruments at Tsinghua University in 2002 and 1997 respectively. He did Post-doctoral research at the University of Calgary for 4 years; and worked as a senior scientist in SiRF Technology Inc. for 3 years. Dr. Niu leads a GNSS/INS group with more than 20 members, including faculties, post doctors and graduate students. The group focus on GNSS/INS deep integration, low-cost navigation sensor fusion, and their applications. Dr. Niu has published more than 70 academic papers and own 9 patents.

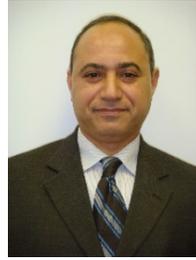

**Naser El-Sheimy** is a Professor and leader of the MMSS Research Group at the University of Calgary and former Chairman of Trusted Positioning Inc., now InvenSense Canada. He holds a Canada Research Chair (CRC) in Mobile Multi-Sensors Geomatics Systems. Dr. El-Sheimy's area of expertise is in the integration of GPS/INS/Imaging sensors for navigation, mapping and GIS applications with special emphasis on mobile mapping systems. He published 2 books, 9 patents, 8 book chapters, and over 400 papers in refereed journals and conferences.

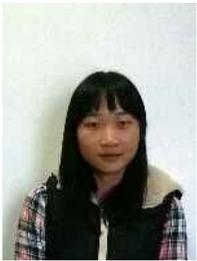

**Qingli Li** is a master student at the School of Geodesy and Geomatics, Wuhan University. She got her bachelor degree of surveying and mapping from the same department in 2014. Her research focuses on GNSS/INS integration technologies.